\documentstyle[twoside,fleqn,espcrc2,amstex,axodraw]{article}
\begin{document}
\input epsf
\newcommand{\mpr}{$\frac{m_\pi}{m_\rho}$}
\newcommand{\plus}{\makebox[15pt][c]{\hspace{-8pt}$+$}}
\newcommand{\minus}{\makebox[15pt][c]{\hspace{-8pt}$-$}}
\newcommand{\err}[2]{\raisebox{0.08em}{\scriptsize {$\;\begin{array}{@{}l@{}}
                          \plus\makebox[0.55em][r]{#1} \\[-0.12em] 
                          \minus\makebox[0.55em][r]{#2} 
                        \end{array}$}}}
\newcommand{\errr}[2]{\raisebox{0.08em}{\scriptsize {$\;\begin{array}{@{}l@{}}
                          \plus\makebox[0.9em][r]{#1} \\[-0.12em] 
                          \minus\makebox[0.9em][r]{#2} 
                        \end{array}$}}}
\newcommand{\er}[2]{\raisebox{0.08em}{\scriptsize {$\;\begin{array}{@{}l@{}}
                          \plus\makebox[0.15em][r]{#1} \\[-0.12em] 
                          \minus\makebox[0.15em][r]{#2} 
                        \end{array}$}}}

\title{Future Perspectives in Lattice Field Theory\thanks{Invited lecture
presented at the XVIIth International Symposium on Lattice Field
Theory, \textit{Lattice 99}, Pisa, Italy, June~28th -- July~3rd 1999.}}
\author{C.~T.~Sachrajda,\\
Dept. of Physics and Astronomy, University of Southampton,\\ 
Southampton, SO17 1BJ, UK 
\vspace{-1.2in}
\begin{flushright}
SHEP 99/19
\end{flushright}
\vspace{0.75in}}

\begin{abstract}
I review some of the contributions which lattice simulations are
likely to make during the next five years or so to the development
of our understanding of particle physics. Particular emphasis is
given to the evaluation of non-perturbative QCD effects in experimentally
measurable amplitudes, and the corresponding extraction of fundamental
parameters. \vspace{-0.15in}
\end{abstract}

\maketitle

\section{Introduction}

I have been asked to talk about the future perspectives of our field,
which I find to be an extremely daunting task. There are 388
participants at this conference, each with his or her own view on how
the subject may or should develop and all I can do is to present one
of these 388 perspectives. The focus of this talk will be on the r\^
ole that lattice simulations are playing now, and will continue to
play in the future in the development of particle physics in general
and in phenomenology in particular. The most important issue is how
well it will be possible to quantify the non-perturbative strong
interaction effects in experimentally measurable quantities. This is
needed in order to be able to deduce fundamental theoretical
information from experimental measurements (e.g. in order to determine
the CKM matrix elements from experimental studies of weak decays).

Much of this talk will be based on the deliberations of the panel set
up last December by ECFA (the European Committee for Future
Accelerators) to advise it on the requirements for high-performance
computing for lattice QCD in Europe and I gratefully acknowledge my
colleagues on the panel for their insights and help~\cite{ecfa}.  The
panel's terms of reference stated that ``the main objective of this
study is to assess the high performance computing resources which will
be required in the coming years by European physicists working in this
field, and to review the scientific opportunities that these resources
would open.''

Most of the effort in improving the precision of lattice computations
of physical quantities will be based on unquenching, the inclusion of
light sea-quarks.  A key ingredient in attempts to estimate the
precision achievable in future simulations is an understanding of the
scaling behaviour of the algorithms used to generate full QCD
configurations (as the quark masses and lattice size and spacing are
varied). I start this talk with a brief review of some recent studies
of this question (section~\ref{sec:resources}). The following sections
contain some of the main physics questions which will be studied using
lattice simulations during the coming years.  The r\^ole of lattice
computations in standard model phenomenology will be considered in
section~\ref{sec:phenomenology}, and will include a discussion of a
selection of physical quantities which are already being computed
frequently and also of other quantities for which lattice calculations
are just beginning or for which we do not (yet?)  understand how to
perform the computations, even in principle.
Sections~\ref{sec:spectroscopy}--\ref{sec:nonqcd} contain brief
discussions of the prospects for lattice simulations of some
quantities in hadron spectroscopy, QCD thermodynamics and non-QCD
physics. Finally, section~\ref{sec:concs} contains some closing
comments.

Perhaps the main excitement at this conference concerned developments
in formulations of chiral fermions on the lattice, and we have had two
very interesting plenary talks on this subject~\cite{chiral}. I will
not discuss these developments because, although their
impact is likely to be very significant, it is still too early to
quantify the effect that the new formulations will have on the physics
discussed below. 

\section{Computing Resources and Lattice Parameters}
\label{sec:resources}

In order to gain some insights into the precision which will be
achievable in the next five years or so, it is clearly necessary to try
to forecast what computing resources might be available and this has
been reviewed by Norman Christ~\cite{christ}. Once it has been
established, that we are aiming at computing power of teraflops, or tens
or even hundreds of teraflops, it is also particularly important to have
a good understanding of the scaling behaviour of the algorithms used to
generate dynamical quark configurations. This is necessary to determine
the lattice sizes and spaces and the values of the quark masses with
which we will be able to simulate. Our understanding of the scaling
behaviour is currently far from complete, and a significant effort will
have to be devoted to the study and development of algorithms.

At this conference S.~G\" usken presented the results of a study based
on the performance of algorithms in the Monte-Carlo Runs (at a lattice
spacing of 0.08 fm, with 2 flavours of Wilson fermions) of the
SESAM/T$\chi$L collaboration~\cite{gusken}. These authors extrapolate
the observed behaviour of their algorithm to larger lattices and lighter
quarks and an illustration of their conclusions is given in
table~\ref{tab:gusken}. The table shows the estimated cost for
generating 100 independent configurations in Tflops-months. 

\begin{table}[t]
\begin{centering}
\begin{tabular}{|c|l|l|l|}
\hline 
\rule[-6pt]{0pt}{17pt}\mpr &   $a=0.10$   &  $a=0.08$&  $a=0.06$  \\
\hline
\multicolumn{4}{|c|}{\rule{0pt}{12pt}2 fm}\\
\hline
\rule{0pt}{12pt}0.60  & 0.037-\bf{0.071} & 
\bf 0.20 & \bf{0.72-1.5} \\
0.50  &   0.073-0.14      &      0.38  & \textbf{1.40}-3.0\\
0.40  &   0.15-0.28       &      0.77  &  2.8-6.0  \\
0.30  &   0.35-0.65       &      1.8   &  6.7-14   \\
\hline
\multicolumn{4}{|c|}{\rule{0pt}{12pt}3 fm}\\
\hline
\rule{0pt}{12pt}0.60   &   \bf 0.24-0.45   &  
\bf 1.2  & \bf  4.6-9.7 \\
0.50     &   \bf 0.46-0.86   &  \bf 2.4  & \bf 8.8-18 \\
0.40     &    0.94-1.8    &  4.9  & \textbf{18}-38  \\
0.30     &   2.2-4.1     &      11   &  42-89  \\
\hline
\multicolumn{4}{|c|}{\rule{0pt}{12pt}4 fm}\\
\hline
\rule{0pt}{12pt}0.60     &   
\bf 0.89-1.7    & \bf 4.6  & \bf  16-36  \\
0.50     &   \bf 1.7-3.2    &  \bf 8.8  & \bf  32-69  \\
0.40    &   \bf 3.5-6.5    &  \bf 18   & \bf  66-141  \\
0.30     &       8.2-15     &      42   &\textbf{152}-332 \\
\hline
\end{tabular} 
\caption{ Estimated costs for generating 100 independent
  configurations in Tflops-months~\protect\cite{gusken}.
  For $a\ne 0.08$ the two
  bounds are given to reflect the uncertainty in the behaviour with the lattice spacing.
  The results from lattices which have an extent of 5 or more pion
  correlation lengths are printed in bold type.}
\end{centering}\vspace{-0.3in}
\label{tab:gusken}\end{table}

Some authors view the estimates in table~\ref{tab:gusken} as being too
optimistic. The CP-PACS collaboration have recently tried to
\textit{estimate the computer time required for a large-scale full QCD
calculation, with the quality of data comparable to that of the present
quenched QCD study on the CP-PACS}~\cite{cppacs}. They use a
renormalisation group improved gauge action and two flavours of
degenerate quarks with the clover fermion action and estimate the
required time to be of the order of 100 TFlops$\cdot$year. For example,
they estimate that they would require 409 days on a 131 TFlops machine
(e.g. on a machine of 4096 PU's with a processing power of 32 GFlops/PU)
in order to obtain 25,000 trajectories on a (3\,fm)$^3$ spatial lattice
with light quarks with masses down to 15\,MeV, for which
$m_\pi/m_\rho\simeq 0.4$ (realized by working on a $48^3\times 96$
lattice with $a=0.067$\,fm).

A similar study is currently underway in the US, being made as part of
the preparation for a proposal to the SSI (Scientific Simulation
Initiative), and the preliminary conclusion is also that the G\"usken et
al. estimate is too optimistic by at least one order of
magnitude~\cite{norman}.

\begin{table}
\begin{center}
\begin{tabular}{ccc}
Authors & Estimate & Action \\ \hline
CP-PACS~\cite{cppacs} & 150 Tflops$\cdot$Years & RGI, Clover\\
GLS~\cite{gusken} & 13 Tflops$\cdot$Years & Wilson\\
Sharpe~\cite{sharpe} & 2.5 Tflops$\cdot$Years & Staggered\\ 
\end{tabular}
\caption{Three estimates of the time required to generate about 1000
independent configurations on a 3\,fm lattice, with lattice spacing
0.067\,fm and $m_\pi/m_\rho=0.4$.}\label{tab:time}
\vspace{-0.3 in}\end{center}\end{table}

In table~\ref{tab:time} I present three estimates of time required to
generate 1000 independent configurations on a 3\,fm lattice, with
lattice spacing 0.067\,fm and $m_\pi/m_\rho=0.4$. This is essentially
the CP-PACS scenario described above. Note that the estimate by Steve
Sharpe is for two flavours of staggered fermions, and experience teaches
us that the time required for Wilson-like fermions is at least a factor
of 10 longer. This discussion demonstrates the importance of further
investigations into the scaling behaviour of algorithms.  The different
groups should start by agreeing on the behaviour in the range of lattice
parameters which they are able to simulate. There will naturally remain
uncertainties when extrapolations are made to larger lattices, smaller
lattice spacings and lighter values for the quark masses. In spite of
the uncertainties, and the obvious expense in computing resources
required to perform full QCD simulations with light quarks, the
estimates above do indicate that the exciting goal of performing
unquenched simulations at the same level of precision as current
quenched ones within the next five years or so is a realistic one.

Norman Christ has reviewed the status and prospects for machines for
lattice simulations~\cite{christ}. For the purposes of this talk I will
simply assume that at least some of the major collaborations will have
access to machines with $O$(10\,TFlops) sustained power within the next
five years or so.

\section{Lattice Phenomenology}
\label{sec:phenomenology}

In this section I will discuss the prospects for quantifying
non-perturbative QCD effects for a variety of physical quantities of
phenomenological interest. For some of these quantities we understand
very well how to perform the computations whereas for many others we
do not. In the former category are many quantities which are well
measured experimentally, and for which computations are used to
demonstrate that lattice systematic uncertainties are under
control. This then gives us confidence in the reliability of
computations of unknown hadronic matrix elements and other quantities
needed to control non-perturbative QCD effects in physical processes.

\begin{figure}[th]
\begin{center}
\begin{picture}(235,125)(0,-20)
\unitlength 1pt
\Line(0,0)(200,0)\ArrowLine(0,0)(140,80)\ArrowLine(140,80)(200,0)
\Text(150,85)[b]{$A=(\bar\rho,\bar\eta)$}\Text(138,72)[t]{$\alpha$}
\Text(20,-10)[t]{$C=(0,0)$}\Text(22,2)[bl]{$\gamma$}
\Text(190,-10)[t]{$B=(1,0)$}\Text(188,2)[br]{$\beta$}
\Text(23,43)[l]{$\bar\rho + i \bar\eta$}
\Text(172,28)[r]{$1-(\bar\rho + i \bar\eta$)}
\end{picture}
\vspace{-0.35 in}
\caption{Unitarity Triangle corresponding to the
relation in eq.(\ref{eq:vckm}).}\label{fig:ut}
\end{center}
\vspace{-0.385 in}
\end{figure}
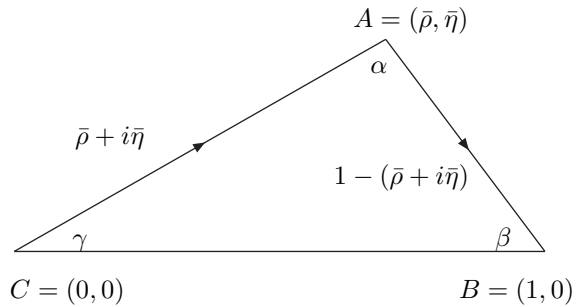
\begin{figure}[t]
\vspace{-0.34in}
\begin{center}
\hbox to\hsize{\hss
\epsfxsize1\hsize\epsffile{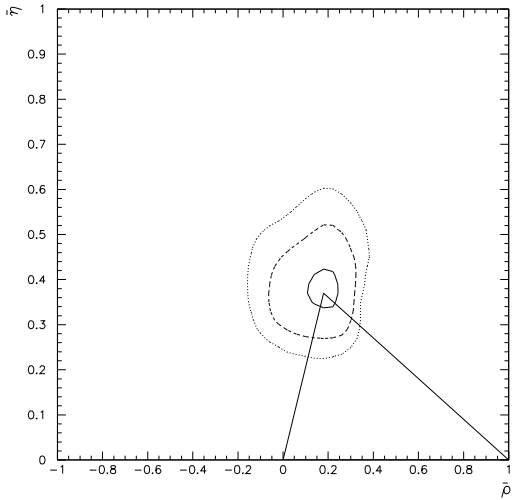}\hss}
\vspace{-0.5 in}
\caption{Currently allowed values of the vertex $A$ of the Unitarity
Triangle~\protect\cite{guido_ut}. The three regions correspond to 5\%,
68\% and 95\% confidence levels.}
\label{fig:now}
\end{center}
\vspace{-0.4in}\end{figure}

It has become conventional to present much standard model phenomenology in
terms of the unitarity triangle. The unitarity of the CKM-matrix 
(written in the Wolfenstein parametrisation~\cite{wolfenstein}),
\begin{equation}
\hspace{-0.3in} \begin{pmatrix} 1-\frac{\lambda^2}{2} &
\lambda & A\lambda^3(\rho-i\eta)\\ -\lambda & 1-\frac{\lambda^2}{2} &
A\lambda^2 \\ A\lambda^3(1-\rho-i\eta) & -A\lambda^2 & 1 
\end{pmatrix}\,,
\label{eq:vckm}\end{equation}
leads to six unitarity relations, including the one most frequently
considered which is shown in Fig.~\ref{fig:ut} (where
$\bar\rho=\rho(1-\lambda^2/2)$ and $\bar\eta=\eta(1-\lambda^2/2)$).
Different physical processes give different loci for the position of the
vertex $A$. In principle from the intersection of these curves we can
determine the position of $A$, of if more than two curves do not
intersect at the same point, we would deduce that effects of
\textit{physics beyond the standard model} are present. In practice,
however, hadronic uncertainties are sufficiently large that we arrive at
a region of allowed positions for $A$. Fig.~\ref{fig:now} contains the
allowed region deduced from one global analysis of current
data~\cite{guido_ut}. The joint aim of the theoretical and experimental
communities is to reduce this region (see below).

The remainder of this section is divided into two parts. In
subsection~\ref{subsec:freq} I discuss some quantities which are
computed frequently in lattice simulations whilst in
subsection~\ref{subsec:infreq} I consider others, for which lattice
simulations are either just beginning or for which some conceptual
progress is needed before such simulations will be possible.

\subsection{Frequently Computed Physical Quantities}
\label{subsec:freq}

\begin{figure}[t]
\vspace{-0.34 in}
\begin{center}
\hbox to\hsize{\hss
\epsfxsize1\hsize\epsffile{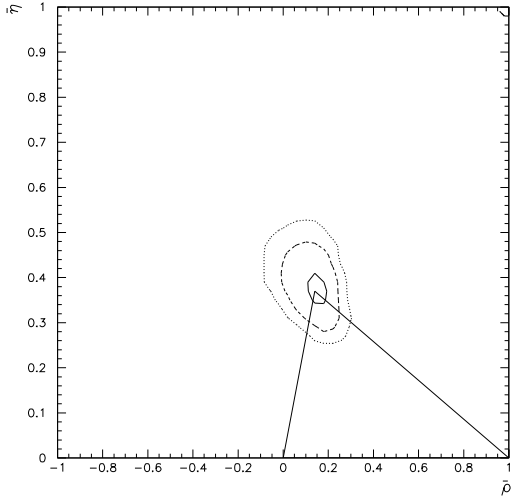}
\hss}
\vspace{-.5 in}
\caption{Scenario for the allowed region for the vertex $A$ of the
Unitarity Triangle using the estimates of future errors as in
table~\ref{tab:errors}~\protect\cite{guido_ut}. The three regions
correspond to 5\%, 68\% and 95\% confidence levels.}
\label{fig:then}
\end{center}
\vspace{-0.3in}\end{figure}

In table~\ref{tab:errors} I present our preliminary estimates of the
uncertainties in current and future results for a selection of physical
quantities and matrix elements which are frequently computed in lattice
simulations~\cite{ecfa}. The estimates in future calculations are based
on being able to generate 1000 configurations at a lattice spacing of
0.08\,fm with $m_\pi/m_\rho=0.4$ and a sketch of how the improved 
precision will reduce the allowed region for $A$ is presented in
fig.~\ref{fig:then} (without assuming any improvement in the 
precision of experimental measurements).
\begin{table}[t]
\begin{center}
\begin{tabular}{|c|c|c|}\hline
quantity & Present & Future \\
& Uncertainty & Uncertainty \\ \hline
$\alpha_s$ & 10\% &2\% \\
$m_{u,d,s,c}$ & 20-25\% & 5\% \\ 
$m_b$ & 5\% & 1\% \\ 
$B_K$ & 15\% & 5\% \\
Other $B$'s & 15-30\% & 5\% \\ 
$f_D,f_{D_s}$ & 15\% & 5\% \\
$f_B, f_{B_s},f_B\sqrt{B_B}$ & 20\% & 7\% \\
$F^{D\to M}(0)$ & 15-20\% & 5\% \\ 
$F^{B\to M}(0)$ & 20-35\% & 7\% \\ 
\hline\end{tabular}
\caption{Estimates of the present and future uncertainties in lattice
calculations for a selection of physical quantities, the strong
coupling constant, quark masses, the $B$-parameter of $K-\bar K$
mixing and other $B$-parameters of four-quark operators, the decay
constants of heavy mesons and heavy$\to$light form factors at zero
momentum transfer.}\label{tab:errors}\end{center}\vspace{-0.3in}
\end{table}

The mass of the strange quark, $m_s$, is an interesting quantity with
which to study the effects of quenching. It has been known for a
considerable time that in quenched calculations one obtains a
different value for $m_s$ depending on whether one uses the $K$- or
$\Phi$-mesons~\cite{mm} (equivalently one cannot reproduce the
physical value of the $J$-parameter, where $J$ is basically the slope
of $m_V^2$ vs $m_{PS}^2$, where $V$ ($PS$) represents vector 
(pseudoscalar) mesons~\cite{cm}). It will therefore be an
important check of our control of full QCD computations to observe
whether this problem is cured. At this conference the CP-PACS
collaboration presented some preliminary results with 2 flavours of
sea quarks, indicating that this may be the case (see the talk by
R.~Mawhinney~\cite{mahw}), and it will be interesting to observe
future developments.

Among the recent interesting analyses in standard model phenomenology,
Parodi, Roudeau and Stocchi have used the measured values of
$|\epsilon_K|, |V_{ub}/V_{cb}|, \Delta m_d$ and the bound on $\Delta
m_s$ to obtain $\bar\rho=0.202\errr{.053}{.059}$ and $\bar\eta=0.340\pm
.035$~\cite{prs} (see also the analysis in ref.~\cite{smele}),
constraining significantly the allowed positions of the vertex $A$ of
the unitarity triangle. These analyses rely on lattice computations of
non-perturbative QCD effects, such as those in $B_K$ (for which the
authors use $B_K=0.86\pm 0.09$) and $f_{B_d}\sqrt{B_{B_d}}$ (for which
they take $210\er{39}{32}$\,MeV), and of course the reliability of the
analyses depends on that of the lattice computations. These authors,
however, then perform an interesting exercise, by not including the
lattice value of each of the hadronic parameters in turn and instead
determining it from the analysis. In this way they find
\begin{eqnarray}
B_K&=& 0.87\errr{0.34}{0.20}\hspace{0.4 in}\textrm{and}\\ 
f_{B_d}\sqrt{B_{B_d}}& =& 223\pm 13\,\textrm{MeV}\ .
\end{eqnarray}
It is interesting to see the parameters which we are used to calculating
in lattice simulations being obtained in this way, albeit relying on
lattice values for the remaining matrix elements and assuming no new
physics.

One of the decay constants, $f_{D_s}$, has been measured directly
by several experiments, which the authors of ref.~\cite{prs} combine
to obtain~\footnote{Note that the Particle Data Group, more cautiously, 
present the spread of values from all experiments for $f_{D_s}$~\cite{pdg}.}
\begin{equation}
f_{D_s}=241 \pm 32\,\textrm{MeV}\ , 
\label{eq:fdsexp}\end{equation}
in excellent agreement with lattice \textit{predictions}, (e.g. in
ref.~\cite{fs} we quoted $f_{D_s}=220\pm30$\, MeV as the average value
from lattice simulations). The experimental result for $f_{D_s}$ is
therefore a significant encouragement to lattice phenomenologists.

Finally in this subsection I consider the form factors for exclusive
heavy\,$\to$\,light semileptonic and radiative decays, and in
particular for $B\to\pi,\rho$ + leptons and $B\to K^*\gamma$
decays. The requirement that the $B$ and final-state light mesons have
small momenta (in order to avoid discretisation errors) implies that
direct lattice calculations only yield these form-factors at large
values of the momentum transfer (i.e.  near the zero-recoil point).
The extrapolation of the lattice results to smaller values of the
momentum transfer leads to an important source of systematic errors,
even though we do have some theoretical guidance from the heavy quark
effective theory (HQET), chiral perturbation theory and from axiomatic
properties of field theories (such as analyticity and unitarity) in
performing these extrapolations~\cite{ukqcdff}. The errors due to the
extrapolation will be significantly reduced by increased computing
power (allowing us to go to smaller lattice spacings and hence to
larger momenta for the mesons). This is an illustration of the fact
that for heavy flavour physics, increased computer power may also be
fruitfully used to perform larger quenched computations.  It is also
to be hoped that there will be progress in understanding the scaling
behaviour of form-factors at small momentum transfers to
make the extrapolations more controllable.

\subsection{Calculations which are Performed Less Frequently:}
\label{subsec:infreq}

The quantities considered in table~\ref{tab:errors} represent an
important set of phenomenologically important parameters, but there
are also many other matrix elements and physical quantities for which
we need to control the non-perturbative QCD effects. I discuss a
selection of these in this subsection.

The major difficulty in lattice phenomenology is that we have no
general method for dealing with multihadronic states and final state
interactions (see for example ref.~\cite{mt}). We can compute amplitudes
for processes with two particles at rest, which, when combined
with chiral perturbation theory proves to be useful for kaon decays
(but not for B-decays). I now consider exclusive nonleptonic
kaon and B-decays in turn.

\subsubsection{$\mathbf{K\to\pi\pi}$ Decays:} $K\to\pi\pi$ decays were
considered in some detail during the early period of lattice
phenomenology. As a result of the difficulties which were encountered
in the evaluation of the corresponding matrix elements and also
because of the development of heavy-quark physics the emphasis of
lattice simulations changed towards quantities such as those in
table~\ref{tab:errors}. More recently the activity in kaon physics
has increased again and at this conference was reviewed by
Y.~Kuramashi~\cite{kuramashi}.  An important difficulty when studying
kaon decays using Wilson-like quarks is to control the chiral
behaviour (and the corresponding subtraction of power divergences in
many cases). The improvement in lattice technology and computing
resources, together with theoretical developments such as
non-perturbative renormalisation~\cite{npr} and new formulations for
lattice fermions, imply that kaon decays should now become a major
area of lattice phenomenology.  The motivation for this is further
underlined by the new measurements of $\epsilon^\prime/\epsilon$ from
the KTEV and NA48 collaborations~\cite{mangano},
\begin{eqnarray}
\frac{\epsilon^\prime}{\epsilon}&=&
(28\pm 3.0\pm 2.8)\,10^{-4}\ \cite{ktev}\\ 
\frac{\epsilon^\prime}{\epsilon}&=&
(18\pm 4.5\pm 5.8)\,10^{-4}\ \cite{na48},
\end{eqnarray}
which are sufficiently large that one might hope to control the
expected partial cancellations between the matrix elements of the
operators which are conventionally called $O_6$ and
$O_8$~\footnote{Since the conference the Riken-BNL-Columbia
collaboration have presented results from a study using domain wall
fermions in which they find surprisingly that these matrix elements
have the same sign and obtain a negative result for
$\epsilon^\prime/\epsilon$ ($\simeq -(12.2\pm 6.8) 10^{-3}$)~\cite{riken}.}.

A quantitative understanding of the $\Delta I=1/2$ rule in kaon decays
is also very desirable and is an important benchmark for lattice QCD
computations (note the interesting result in ref.~\cite{pekurovsky},
which however has large errors bars, and, as is frequently the case
for staggered fermions is plagued by very large perturbative
corrections in the matching factors).

\subsubsection{$\mathbf{B\to MM}$\,Decays:} There is a flood of data
becoming available on exclusive decays of $B$-mesons into two
light-mesons, potentially giving fundamental information about the CKM
Matrix and CP-violation. This flow of results will increase still
further as the $b$-factories and other approved experiments start taking
data. Again the difficulty is in controlling the non-pertubative QCD
effects. All the problems of non-leptonic kaon decays are present again,
but in addition, the use of chiral perturbation theory is not applicable
in $B$-decays. New ideas are needed urgently and it may be that it is
necessary to combine lattice calculations with models in order to make
progress (see for example ref.~\cite{ciuchini}). Any new ideas can also
be tested on the huge amount of experimental data which already exists
for nonleptonic charm decays.

The difficulty in controlling hadronic effects in nonleptonic
$B$-decays serves to underline the beauty of the one
\textit{solid-gold} process of $B\to J/\Psi K_s$, which is essentially
free of these uncertainties and which will provide an accurate
determination of the angle $\beta$ at the $b$-factories~\footnote{The
recent measurement from the CDF collaboration gives
$\sin(2\beta)=0.79\err{0.41}{0.44}$~\cite{fnalbeta}.}.

\subsubsection{$\mathbf{B}$-Lifetimes:} Lattice calculations of rates for
inclusive non-leptonic decays are beginning, and are likely to make an
important contribution to standard model phenomenology (see the review
talk by Hashimoto~\cite{hashimoto}). To leading order in the
heavy-quark mass all beauty hadrons are predicted to have the same
lifetimes, and the corrections to this prediction can be calculated
using an expansion in inverse powers of the mass of the
$b$-quark~\cite{bsuv} (see also refs~\cite{suv} and \cite{mn} for
reviews and references to the original literature). For the ratio of
lifetimes of the $\Lambda_b$-baryon and the $B$-meson the expansion
leads to the prediction:
\begin{equation}
\frac{\tau(\Lambda_b)}{\tau(B_d)}= 1 + 0 - 2\% +O(\frac{1}{m_b^3})\ ,
\label{eq:taus}\end{equation}
where the $0$ on the right-hand side indicates that there are no
operators of dimension 4 which can contribute and the -2\% is an
estimate of the contribution of dimension 5 operators obtained by
comparing the spectroscopy of charmed and beauty mesons and baryons. In
view of eq.~(\ref{eq:taus}) the experimental measurement,
$\tau(\Lambda_b)/\tau(B_d)=0.78(4)$ is very puzzling. One possibility is
that spectator effects (i.e. effects in which the light quark
constituent of the hadrons participate directly in the weak decay
process), which in the heavy quark expansion appear at $O(1/m_b^3)$ but
which have a phase-space enhancement~\cite{ns}, are sufficiently large
to account for the experimental result. This can be checked in lattice
calculations, by computing the matrix elements of the corresponding
four-quark operators. A recent exploratory calculation shows that
spectator effects are indeed significant, and give a contribution to the
right-hand side of eq.(\ref{eq:taus}) of -(6-10)\%~\cite{dps,dpms}. This
calculation can readily be improved, and it is important to do so to
learn about the practicability of using the heavy quark expansion for
predictions of inclusive nonleptonic decays.

A related quantity is $\Delta\Gamma_s/\Gamma_s$ for which the first
lattice results where presented at this conference by the Hiroshima
group~\cite{hiroshima}.

\subsubsection{Lightcone wavefunctions:} Light-cone wave functions contain the
non-perturbative QCD effects in hard exclusive processes (such as the
form-factors at large momentum transfers~\cite{bl,er} or nonleptonic
$B$-decays~\cite{bbns}) and again lattice computations can make a
significant contribution to the evaluation of these effects. To my mind,
however, there has been insufficient effort in this field. In addition
to calculating the moments of these wave-functions in the traditional
way, it will also be interesting to explore and develop new suggestions
for computing the wave-functions themselves~\cite{guido}.

The main aim of this section was to stress the wide range of QCD
phenomenology to which lattice computations can make a significant
impact. Of course there are many examples in addition to those presented
here, such as deep inelastic structure functions~\cite{roberto}, proton
decay amplitudes~\cite{kuramashi}, shape functions for inclusive
heavy$\to$light decays~\cite{guido2}, the electric dipole moment of the
neutron etc.

\section{Hadronic Spectroscopy}
\label{sec:spectroscopy}

Hadronic spectroscopy continues to be a benchmark of central importance
for lattice computations and the emphasis is now clearly on performing
unquenched calculations~(see the comprehensive review by Bob
Mahwinney~\cite{mahw}). In order to extrapolate the results obtained
directly from lattice simulations to those corresponding to physical $u$
and $d$-quark masses, guidance is needed from chiral perturbation
theory. Estimates of how small the masses in simulations need to be in
order for chiral perturbation theory to lead to reliable extrapolations
are typically in the range corresponding to $m_\pi/m_\rho\simeq
0.3$--0.4 (see ref.~\cite{sharpe} and references therein) and we have
seen in section~\ref{sec:resources} that this is at the limit of what we
might hope to achieve in the next few years. A particularly important
milestone will have been reached when we are able to observe and control
$\rho\to\pi\pi$ decays, which given the $P$-wave suppression of the
decay-rates will take some time.

\begin{table}
\begin{center}
\begin{tabular}{cc}
Quantity & Uncertainty ~ 2003-5\\ \hline
$m(\eta^\prime)-m(\eta)$ & 30\,MeV\\
$0^{++}$ Mixing & 10\% \\
$m(1^{-+})$ & 20\,MeV\\
$m(H)$&20\,MeV
\end{tabular}
\caption{Guesstimates of the likely uncertainties which might be achieved
for a number of key spectroscopic quantities by 2003-5 or so.}
\end{center}
\label{tab:spect}\vspace{-0.3in}\end{table}

In table~\ref{tab:spect} I have presented guesstimates of the relative
precision which might be achieved for some key quantities in
spectroscopy during the next three to five years or so~\cite{ecfa}. A
significant effort will be required to get a reasonable handle on all
of these quantities.  Samples of very large numbers of configurations
($10^4$ or more) are required to get good signals for flavour singlet
mesons.  For the exotic $1^{-+}$ state Doug~Toussaint reminded us that
the lattice result of 1.9\,GeV should be compared with the
experimental value of 1.4\,GeV, so that we need to study mixing with
meson-meson channels~\cite{toussaint}. Note that the current
theoretical uncertainty on the mass of the $1^{-+}$ meson is about
200\,MeV.

It will also be interesting and important to start to do ``nuclear
physics" with lattice simulations. These studies will presumably start
with the simplest systems, such as the deuteron or the H-dibaryon
(bound-state of two $\Lambda$-baryons). For these computations to be
possible large volumes will be required, but, in addition, the fact
that the binding energies are small, $O(10\,$MeV), implies that again,
high statistics will be needed.

\section{QCD Thermodynamics}

The commissioning of the RHIC accelerator at Brookhaven and the approval
of the Alice experiment at CERN means that lattice thermodynamics will
have a major phenomenological r\^ ole to play as well as a theoretical
one. At this conference Fritjof Karsch reminded us that the key features
of the quenched theory are well understood (the existence of a
first-order phase transition, the equation of state, a critical
temperature of about 270\,MeV) and that here also the attention is
focused on simulations with dynamical quarks. The NUPECC Report on
Computational Nuclear Physics concluded that in order to study QCD
thermodynamics in simulations with light quarks, a lattice spacing of
about 0.1\,fm and a volume of 100\,fm$^3$, about 10\,Tflops$\cdot$years
of computing effort are required. It appears that the effects of
quenching are very significant, for example the critical temperature
decreases substantially as quark loops are included ($T_c\to170$\,MeV
for two sea-quark flavours). In simulations with computing power in the
10 Teraflops range, it is expected that it will be possible to determine
the critical temperature with a 5\% accuracy.

\begin{figure}
\begin{center}
\hbox to\hsize{\hss
\epsfxsize1.0\hsize\epsffile{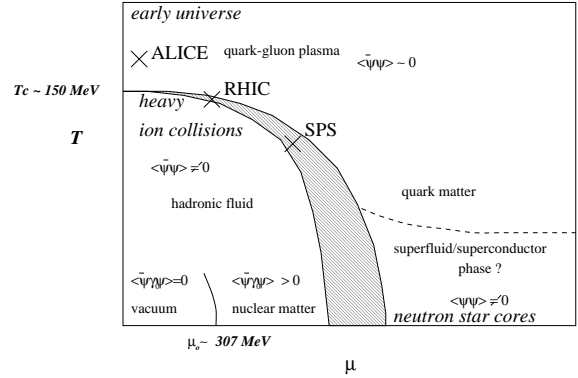}
\hss}
\vspace{-0.35 in}
\caption[]{Schematic Phase Diagram for QCD.}
\label{fig:superc}
\end{center}
\vspace{-0.5in}\end{figure}

A schematic diagram of the expected phase structure of QCD in the
$(T,\mu)$ plane is presented in figure~\ref{fig:superc} (prepared by
Simon Hands), where $\mu$ is the chemical potential. Verification that
the structure is indeed as expected, and in particular investigations
of the fascinating scenario of a superconducting phase (or
phases) of QCD at high $\mu$ (for a recent review and references to the
original literature see ref.~\cite{wilczek}), represents a major
challenge for the lattice community (see the review by Shuryak at this
conference~\cite{shuryak}). At non-zero $\mu$, the action is complex,
and conceptual progress is needed to develop reliable lattice
calculations. It is, however, possible to study other quantum field
theories with similar expected structures (but for which the action is
real) and over the next few years we can look forward to the insights
which these simulations will yield.

\section{Non-QCD Physics}
\label{sec:nonqcd}

In this talk I have concentrated on lattice QCD, but we should bear in
mind the many application of lattice field theory to other areas of
particle physics, some of which we have heard about at this conference.
I will now briefly mention applications to the electroweak phase
transition and to supersymmetric gauge theories; other applications
include the study of phase structures in general (e.g. in QED), other
applications in electroweak physics (strong Yukawa sectors, vacuum
stability and mass bounds, heavy and strongly interacting Higgs bosons),
and studies in quantum gravity. For many non-QCD applications the
developments in new formulations of lattice fermions are likely to play
a key r\^ole.

\subsection{The Electroweak Phase Transition}\label{subsec:ewpt}

The motivation for lattice studies of the electroweak phase transition
is to gain an understanding of the baryon asymmetry of the universe and
this subject has been comprehensively reviewed at this conference by
Fodor~\cite{fodor}. For Higgs masses greater than about 60\,GeV,
infrared problems invalidate the use of perturbation theory, requiring
the use of lattice simulations. The phase diagram determined from
lattice simulations shows a line of 1st-order phase transitions in the
$(T_c,m_H)$ plane, terminating at a 2nd order end-point at $m_H\simeq
72$\,GeV. In order for sphaleron production to be the mechanism for the
generation of the observed baryon asymetry, as expected in the standard
model, there must be a strong first order transition, and hence the mass
of the Higgs boson must be less than 72\,GeV, in contradiction with the
experimental bound of 95\,GeV from LEP~\footnote{72\,GeV is the position
of the 2nd-order end-point. For a sufficiently strong transition the
bound is even stronger, $m_H< 40\,$GeV, even more clearly in
contradiction with experiment.}. It would be reassuring to include the
fermions explicitly in these simulations (which would require major
computing resources) but already these results present significant
evidence that physics beyond the standard model is required.

In the minimal supersymmetric standard model the bound is weaker,
$M_H<110\,$GeV, just above the experimental limit from LEP.

\subsection{Supersymmetry}

Much of the phenomenology of physics ``beyond the standard model" is
based on supersymmetric field theories. These theories, with their high
degree of symmetry, have extremely interesting non-perturbative
phenomena (an outstanding example is Seiberg-Witten duality). The
subsequent developments have dramatically deepened our understanding of
non-perturbative features of gauge theories, although it is not yet
clear what the implications for non-supersymmetric theories such as QCD
are. This makes the potential investigations of supersymmetric theories
using lattice simulations very exciting.

Lattice studies of supersymmetric gauge theories are a formidable
challenge and it is natural to start with simpler field theories such as
$N=1$ supersymmetric Yang-Mills theory and to investigate details of the
expected non-perturbative features of confinement and spontaneous chiral
symmetry breaking. The computational requirements are similar to those
for full QCD and current studies are in their infancy (teaching us for
example that the $SU(2)$ theory has spontaneous discrete chiral symmetry
breaking caused by a gaugino condensate). Simulations with small fermion
masses are expensive (the supersymmetric limit has vanishing gluino
mass), and computing power of the order of a Teraflop sustained would
enable most of the important theoretical questions to be answered (mass
spectrum, the SUSY potential, Ward identities). The $N=2$ SUSY theory
has a considerably more complicated parameter space and so a large
amount of effort will be required to explore its phase diagram and so no
final results can be expected soon. It remains a very important
long-term goal.

\section{Conclusions}
\label{sec:concs}

It is my belief that in order for our community to continue to receive
support in general, and for state-of-the-art computing facilities in
particular, we have to be perceived as having a definite r\^ole in the
development of particle physics. This is certainly the case at present
and in this talk I have tried to illustrate the wide range of physical
quantities and processes for which lattice results are being used
extensively to quantify non-perturbative effects (see also the talk by
Mangano~\cite{mangano}).  In many cases the leading obstacle to
determing standard model parameters or other fundamental information
from experimental measurements, is not due to experimental
difficulties but to our inability to control the lattice systematic
uncertainties as precisely as we would like. In discussing how our
computations will improve in the future, it is easy to say that we
will do the best that can be done with the resources which will be
available. This is probably true, but in my view insufficient, and I
believe that we also need to try and predict the ``physics-reach'' of
the next generation (or two) of simulations. No doubt these
predictions will have to be revised with time, but we need our
theoretical and experimental colleagues to have a realistic picture of
our expectations, which are considerable but nevertheless limited.
Together with my colleagues on the ECFA panel we have tried to think
about the prospects and strategies for the next five years, and some
of our preliminary estimates can be found above. I hope that this will
be a useful contribution to a debate within the community about the
future of lattice field theory.

\subsection*{Acknowledgements} I warmly thank my colleagues on the  ECFA
panel for the many discussions on the topics discussed in this talk. I
am grateful to Y.~Iwasaki for providing me with the information about
the CP-PACS study of the behaviour of their dynamical fermions
algorithm, and to Z.~Fodor, S.~Hands (who provided me with
figure~\ref{fig:superc}, J.~Kuti, I.~Montvay and S.~Sharpe for
instructive discussions. This work was supported by PPARC grants
GR/L29927 and GR/L56329.

\end{document}